\documentclass{ws06-procs}

\begin{document}

\title{Strong lensing, dark matter and $H_{0}$ estimate}

\author{C. Tortora}

\address{Osservatorio Astronomico di Capodimonte
Salita Moiariello, 16, 80131 - Napoli - Italy\\E-mail:
ctortora@na.astro.it}


\maketitle

\abstracts{Gravitational lensing represents a powerful tool to
estimate the cosmological parameters and the distribution of dark
matter. I will describe the main observable quantities,
concentrating on strong lensing, that manifests its effect through
the formation of spectacular events, like multiple quasars,
Einstein rings and arcs in clusters of galaxies. In events where a
quasar is lensed by an intervening galaxy, it is possible to give
an estimate of the Hubble constant $H_0$, by choosing a mass
density model for the lens, thus allowing to constrain the dark
matter content.}

\section{Introduction: historical remarks}

Gravitational lensing (GL), born as a theoretical prediction, has
became a powerful tool to explore the universe. It concerns all
those phenomena linked to the deflection of light of faraway
sources due to the presence of an intervening astrophysical body,
named `lens'. The concept of light deflection is also present in
the Newtonian physics: in fact during 1700-1800 some scientists,
like Newton, Cavendish and Soldner studied this problem giving an
estimate of this deflection.

In the beginning of the last century, Einstein elaborated its
revolutionary theory of General Relativity, allowing him to give
the correct prediction for the deflection angle of a point mass
lens, given by
\begin{equation}
\hat{\alpha}=\frac{4 G M}{c^{2} \, r},\label{eq:alfa_Einstein}
\end{equation}
where $M$ is the mass of lens and $r$ is the distance at which the
ray passes near the lens. Eq. (\ref{eq:alfa_Einstein}) is twice of
the classical prediction. Observing the change in the position of
stars on the limb of the sun during an eclipse in 1919, Sir A.
Eddington was able to verify the prediction of Einstein, thus
providing the first experimental proof of General Relativity .

In 1934 Zwicky was the first to consider that also galaxies would
be able to act as lenses, increasing the possibility of observing
these phenomena also on larger scales. In 1964 Refsdal analyzed
the possibility to use GL as a tool to study cosmological
parameters and in particular to estimate Hubble constant.

In 1979 Walsh, Carswell e Weymann~\cite{SEF} observed the first
lens, a couple of nearby and very similar quasars, images of the
same source, the double quasar Q 0957+561. In the next two decades
other events were observed: many other multiple lensed quasars,
arcs and arclets in clusters of galaxies, Einstein rings,
extragalactic and galactic microlensing, weak lensing, etc.
etc.~\cite{SEF}

\section{Basics of gravitational lensing}
GL takes place when a lens (a star, a galaxy or a cluster of
galaxies) crosses (or is near to) the line of sight to a distant
source. Under the ``weak field approximation'', the lens only
perturbs ``little'' the space-time metric, which can be written
as~\cite{SEF,Saas_fee,Straumann}
\begin{equation}
d s^{2}= \bigg (1+ \frac{2 U}{c^{2}}\bigg) d t^{2} - \bigg (1-
\frac{2 U}{c^{2}}\bigg) d \textbf{x}^{2}\label{eq:metric}
\end{equation}
where $(t,\textbf{x})$ are time and 3-D space coordinates, and
$U(\textbf{x})$ the gravitational potential. We exclude ``strong
fields'', like those generated by black hole and neutron stars,
that need more a complex analysis.

We can distinguish three regimes

\begin{itemize}
\item \emph{strong lensing} that consists in the formation of multiple
(magnified) images of the source (lensed quasars, Einstein rings
and arcs in clusters of galaxies),
\item \emph{microlensing} as a case of strong lensing, when the
images are blended and undistinguishable (extragalactic and
galactic microlensing),
\item \emph{weak lensing} characterized only by a deformation and a light
amplification of the images without formation of different images;
this regime is practically present in every astronomical image.
\end{itemize}

We will analyze the first regime, mainly describing the
observables that characterize it and concentrating on the
spectacular events it can generate. We can identify three main
observable phenomena:
\begin{enumerate}
\item formation of images,
\item (de)amplification of images,
\item time delay among the images.
\end{enumerate}

\subsection{Deflection angle and lens equation}
\begin{figure}[ht]
%\epsfxsize=10cm   %width of figure - will enlarge/reduce the figures
%\epsfbox{fig3.eps}
%\figurebox{4cm}{5cm}{} %to have a box alone
\centerline{\epsfxsize=1.5in\epsfbox{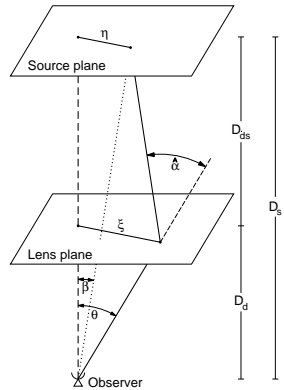}}
\caption{Typical lens configuration of a lensing events. We
observe the \emph{lens} and \emph{source planes}. The long dashed
line represents the line of sight observer-lens, the vector
$\vec{\eta}$ (and the corresponding angle $\vec{\beta}$) specifies
the position of the source on the relative plane, and the vector
$\vec{\xi}$ (and the relative angle $\vec{\theta}$) the position
of the generic images on the lens plane. Crucial ingredients in
such a configuration are the distances, $D_{s}$, $D_{d}$ and
$D_{ds}$. In cosmological applications, angular diameter distances
are used, depending on the redshift $z$ of the considered object
and on the cosmological model
parameters.\label{fig:lens_configuration}}
\end{figure}

The prediction of Einstein regarded the deflection angle of a star
of mass $M$. In more general situations, when the lens is a galaxy
or a cluster of galaxies, this expression is more complex: the
deflection will depend on the mass distribution of the lens. Weak
field approximation allows us to use Eq. (\ref{eq:metric}) and to
manage small angle~\cite{SEF} (in common situations, the angular
separations of the images is $< 30''$). Therefore, we can analyze
a lensed events in a thin cone around the line of sight
observer-lens, studying the phenomena in two plane builded on the
lens (\emph{lens plane}) and source (\emph{source plane}). In
addition, the lens can be considered ``thin'' (since the extension
along the line of sight is small respect to the transversal
dimensions), and therefore all we need to know are the quantities
projected on the lens plane, first of all the surface mass density
of the lens $\Sigma(\xi)$ (see, for instance, Fig.
\ref{fig:lens_configuration}).

\noindent For a generic surface mass density $\Sigma(\vec{\xi})$,
the \emph{lensing angle} is given by
\begin{equation}
\vec{\hat{\alpha}}(\vec{\xi})\equiv\frac{4G}{c^2}{\int}\frac{\vec{\xi}-\vec{\xi}'}{{\vert
\vec{\xi}-\vec{\xi}'\vert}^2}\Sigma(\vec{\xi}')d^{2}\xi',\label{eqlentexi}
\end{equation}
We can derive a two dimensional relation between the positions of
images and source, the so called \emph{lens equation} or
\emph{lens mapping}
\begin{equation}
\vec{\beta}=\vec{\theta}-\vec{\alpha}(\vec{\theta})\label{eq:lens_equation}
\end{equation}
where
$\vec{\alpha}(\theta)=\frac{D_{ds}}{D_{s}}\vec{\hat{\alpha}}(\theta)$
is the \emph{adimensional lensing angle}. Given the source
position angle $\vec{\beta}$, Eq. (\ref{eq:lens_equation}) is a
non linear equation of the image position angle $\vec{\theta}$,
thus a generic lens distribution generates more than 1 image.

\subsection{Lensing magnification} A lens also acts as a sort of a
telescope, magnifying the sources. In addition to a deformation of
the path of a light ray, the lens modifies the transversal section
of a light bundle, enlarging or reducing its sectional area and
changing its form. The lens does not affect light intensity, and
thus the amplification of an image only depends on the angles
subtended by the image itself and source.

\subsection{Time delay}
The images of a lensed source are characterized by a time delay,
i.e. in one of the images the lensed object will be observed
before than the other ones. The time delay between a generic path
from source to observer and the unlensed ray is given
by~\cite{SEF,Saas_fee,Straumann}
\begin{equation}
\Delta t (\vec{\beta},
\vec{\theta})=\frac{1+z_{d}}{c}\frac{D_{d}D_{s}}{D_{ds}}\bigg (
\frac{1}{2}(\vec{\theta}-\vec{\beta})^{2}- \psi(\vec{\theta})
\bigg), \label{eq:time_delay}
\end{equation}
where $z_{d}$ is the lens redshift and $\psi$ is the so called
\emph{lensing potential}, linked to the adimensional lensing angle
$\vec{\alpha}$ by means of the gradient
$\vec{\alpha}=\vec{\nabla}_{\theta} \psi$. It is clear the
contribution of two effects: a \emph{geometric} term, due to the
different paths traveled respect to the unlensed straight line
from the source to the observer, a \emph{gravitational} one, due
to the effect of gravitational potential of the lensing mass.

If the source has a luminosity variable with the time (this is the
case of some quasars), different features in the light curve can
be revealed delayed in the different images and the relative time
lag can be measured. Thus, while $\Delta t$ is not measurable,
since we do not see the unlensed source, it is possible to measure
the quantity
\begin{equation}
\Delta t_{ij}=\Delta t(\vec{\beta},\vec{\theta}_{j}) - \Delta
t(\vec{\beta},\vec{\theta}_{i}),\label{eq:time_delay_ij}
\end{equation}
i.e., the time delay between the $j^{th}$ and $i^{th}$ images in a
lensing event.

\begin{figure}[ht]
%\epsfxsize=10cm   %width of figure - will enlarge/reduce the figures
%\epsfbox{fig3.eps}
%\figurebox{4cm}{5cm}{} %to have a box alone
\centerline{\epsfxsize=3.4in\epsfbox{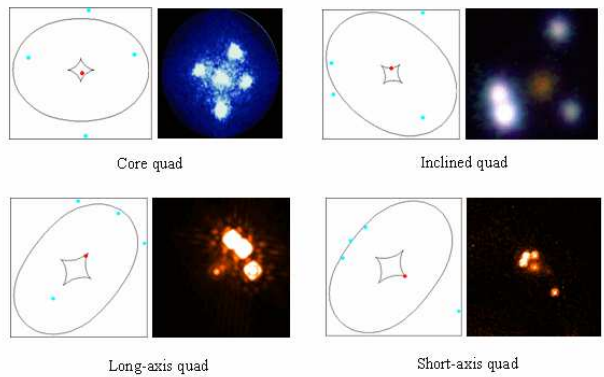}} \caption{Image
configurations in quadruply lensed events. For the double quasars
are possible two configurations that we do not report
here.\label{fig:quads}}
\end{figure}

\subsection{Some details about image formation}

It is possible to classify the images in a simple way. We
distinguish two kinds of images: \emph{ordinary} and
\emph{critical} ones. Ordinary images occur at the points where
the lens equation is verified and where the amplification is
finite. On the contrary, for some critical points the lens mapping
is not invertible and the amplification is infinite. These points
form on the lens plane closed curves, named \emph{critical
curves}, and if we map them to the source plane we get other
closed curves, named \emph{caustics}. We can distinguish two kinds
of critical curves: radial and tangential ones, but the second
ones are crucial in our case to analyze the formation of images.
Limiting the attention to the case of a single lens: for
\emph{spherical mass distribution} the critical curves are
circular, the tangential caustic is point-like and the radial one
is circular, in this case, in addition to a central image, two
images form; for \emph{elliptical mass distribution} the critical
curves are not circular, the tangential caustic becomes an astroid
with folds and cuspids (see Fig. \ref{fig:quads}) and in addition
to an eventual central image, we can have 2 or 4 images.

In particular, when a source approaches the astroid from the inner
side, two images fuse and disappear, thus passing from a 4-images
to a 2-images configuration. In typical lensing events, when the
source is near a fold, we observe two nearby images, one on a side
of the critical curves and another to the other side (see inclined
quad in Fig. \ref{fig:quads}), on the contrary, when the source is
near a cusp, 3 images form near the critical curves (see lower
panels in Fig. \ref{fig:quads}). These images are magnified
respect to the other ones, as can be easily verified in real cases
by an inspections of the quasars in Fig. \ref{fig:quads}. More
complex configurations with a higher number of images are possible
when more than one lens are considered.

\section{Ingredients to shape lensed quasars}
In order to analyze a lensing event, we need to know some
information about lens galaxy and quasar. We need the image
positions respect to the primary lens galaxy, the flux ratios and
time delays, the lens and source redshift $z_{d}$ and $z_{s}$ and
in addition other information about the galaxy (like the luminous
profile, the velocity dispersion profile, etc.). The quantities to
be estimated are source position, the cosmological parameters
(density parameters and $H_0$) and the lens
model~\cite{SEF,Saas_fee}.

Time delay in  Eq. (\ref{eq:time_delay}) can assume the factorized
expression
\begin{equation}
\Delta t_{ij}=
\frac{1}{H_{0}}T(\textrm{distances},z_{d},z_{s})f(\textrm{lens
model},\theta_{i},\theta_{j},z_{d},z_{s})
\end{equation}
where $T$ contains the contribution of the cosmology ($T \to 0$
when is $z_{i} \to 0$ or we consider an Euclidean universe) and
$f$ crucially depends on the lens model. The function $T$ is fixed
assuming some fiducial values for the cosmological parameters,
$H_0$ and the lens model remain to be fitted. In many cases (if we
use standard models of Universe), the uncertainties introduced by
fixing the function $T$ amounts to some percentages. Therefore,
using the constraints above (comprising the time delays), we have
to choice a form for the lens model and fit it to these
observables.

\section{Lens model}

When possible (i.e., when along the line of sight we observe a
single galaxy, with other objects enough faraway from it) we have
to choose a form for the potential of the primary lens and shape
the contribution of other galaxies or external groups of galaxies
with an external shear.

The main galaxy needs a full modeling with the choice of a
particular galaxy model. We can distinguish two classes of models,
linked to the kind of matter we are considering, luminous or dark
matter
\begin{itemize}
\item \emph{Luminous matter}. The stars in galaxy emit a lot of
luminous radiation. Thus, one measures the light profile of galaxy
as a function of the radius from its center. The most credited
models are the de Vaucouleurs law, or other profiles from
Hernquist, Jaffe, Dehnen, Hubble, etc. In order to obtain the mass
profile we use the assumption of a constant M/L ratio, reasonable
since in this case all the matter exists is observed by its
emitted light.
\item \emph{Dark matter}. Galaxies and cluster of galaxies seem to be filled by
a huge quantity of an unseen matter (dark matter). Different
approaches can be chosen to describe dark matter profiles,
starting from theoretical models (singular isothermal sphere,
models starting from distribution function, etc.), models from
simulations (NFW model) or empirical analysis (Sersic model).
\end{itemize}

The external perturbations is taken into account by means of the
so called shear, developing $\psi$ to the $2^{nd}$ order
\begin{equation}
\psi_{\gamma}(r,\theta)= - \frac{1}{2}\gamma \ r^{2} \cos 2
(\theta - \theta_{\gamma}),
\end{equation}
where $\gamma$ is the shear amplitude and $\theta_{\gamma}$ fixes
the direction where is the perturbation.

\section{Estimate of Hubble constant}
$H_{0}$ is a main parameter in cosmology, crucially influencing
the dynamics of the universe. GL is a tool to estimate
cosmological parameters and particularly $H_0$, complementary to
other methods like cosmic distance ladder, anisotropy of the
cosmic microwave background radiation, the Sunyaev-Zel'dovich
effect, etc.

\begin{figure}[ht]
%\epsfxsize=10cm   %width of figure - will enlarge/reduce the figures
%\epsfbox{fig3.eps}
%\figurebox{4cm}{5cm}{} %to have a box alone
\centerline{\epsfxsize=2.9in\epsfbox{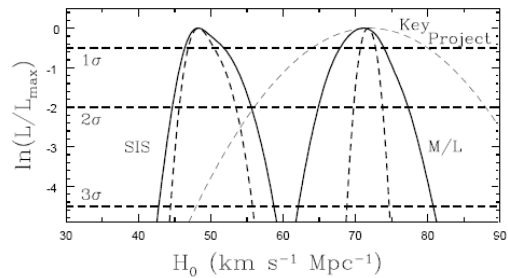}}
\caption{Joint $H_0$ likelihood distribution using 4 lenses and
assuming a SIS (left curves) and a constant $M/L$ model (right
curves) and the results from HST Key project (light dashed
curves).\label{fig:kochanek_saas_fee}}
\end{figure}

GL has many advantages respect to other methods, allowing to
avoid, for example, the propagation of the uncertainties we can
observe in the application of the cosmic distance ladder, but the
choice of lens model remains a huge source of uncertainty. Two
results discussed in literature concern the fact that the same
lens configuration is well reproduced by different classes of
models, and the corresponding dependence of estimated $H_0$ from
the chosen lens model. In particular, constant $M/L$ profiles
systematically furnish higher values of $H_0$ than those given by
profiles with dark matter (like, in the simplest case, the SIS).
See Fig. \ref{fig:kochanek_saas_fee}, where we see the estimated
value from HST Key project is not consistent within $2 \sigma$
with the lensing estimate using a SIS profile, but in agreement
with the estimates using constant M/L profiles~\cite{Saas_fee}.

We verified this result, using more complex dark matter models and
two different constant $M/L$ profiles~\cite{Tortora}. The first
models are fixed by choosing a simple separable form for the
lensing potential
\begin{equation}
\psi(r, \theta)=F(\theta, q, \theta_{q})
r^{\alpha},\label{eq:ell_pot}
\end{equation}
where the function $F$ specifies the angular dependence, $(q,
\theta_{q})$ are the the axial ratio and the angle of the
orientation of lensing potential, and $\alpha$ is a slope
parameter. The constant M/L models are the de Vaucouleurs and
Hubble profiles. The degeneracy and the interesting results above
are evident in our work, also being our estimates affected by huge
uncertainties due to our method and the degeneracies existing
among the lens parameters and $H_0$ itself. Finally, marginalizing
over the results from different models and over two different
lensed quasars we obtain
\begin{equation}
H_{0}=56 \pm 23 \textrm{ km} \ \textrm{s}^{-1} \ \textrm{Mpc}^{-1}
\end{equation}

Using the model in Eq. (\ref{eq:ell_pot}), leaving the function F
unspecified and suitably manipulating the lens and time delay
equations, it is possible to model at the same time different
lensed events that share the same $H_0$. Thus, we can obtain an
estimate of $H_0$ modeling different lensed quasars all at once.
Using 2 lensed quasars, the results is~\cite{Tortora}
\begin{equation}
H_{0}=49^{+6}_{-11} \textrm{ km} \ \textrm{s}^{-1} \
\textrm{Mpc}^{-1}
\end{equation}

\section{A general choice for the lens model}
As remarked above, the choice of lens model remains the main
source of uncertainties in the estimate of $H_0$ by means of
lensed quasars. Since we do not know what is the `right' profile
of galaxy, we can follow two different routes.

The first one was previously discussed and consists to do a
\emph{marginalization over different models} in order to obtain an
estimate of $H_0$ not depending on the particular choice of model,
but in some sense weighing the single estimates obtained using
each model with its uncertainty.

The second one consists to assume a \emph{very general form for
the lens model}, able to reproduce different behaviors, spanning
from $M/L$ profile (with declining rotation curves) to dark matter
profile (with flat rotation curves). We start from this general
requirement and thus choose a double power law expression for the
$M/L$ ratio~\cite{Tortora_varmtol}
\begin{equation}
\Upsilon(r)=\frac{M(r)}{L(r)}=\Upsilon_{0}\bigg (\frac{r}{r_{0}}
\bigg)^{\alpha} \bigg (1+\frac{r}{r_{0}}
\bigg)^{\beta},\label{eq:MtoL}
\end{equation}
where $M(r)$ and $L(r)$ are the mass and luminosity within $r$,
$\Upsilon_0$ a strength parameter, $r_0$ a characteristic radius
and $\alpha$ and $\beta$ are two slope parameters. In particular,
$\alpha$ determines the trend for $r \to 0$ and the sum $\alpha +
\beta$ the trend for $r \to \infty$. From Eq. (\ref{eq:MtoL}) it
is simple to obtain the mass density profile, and dynamical and
lensing properties, if we assume an expression for the light
profile (See Ref.~\cite{Tortora_varmtol} for more details).

%This general choice allows to analyze a sample of lensed events
%(adding information from dynamics) in a homogeneous way, without a
%choice of a fixed models and to select the different lenses in
%according to the estimated values of the lens parameters and
%describe the content of dark matter.

The link between $H_0$ and mass profile is very interesting and
can be useful to reduce the degeneracies existing in lensing
modeling. However, GL remaining a useful and powerful tool to
determine $H_0$, it could be more advantageous to fix $H_0$ to a
reference value determined by means of other methods, and try to
obtain information about mass profile of lens galaxies and to
determine their content of dark matter. In this direction, the
choice of a general class of galaxy models like those in Eq.
(\ref{eq:MtoL}) has many advantages we outline below, giving the
possibility to analyze many different questions
\begin{itemize}
\item many relative dynamics and lensing quantities assume analytical
expressions
\item since it is based on a general expression for the M/L ratio, can allow
to homogeneously analyze a huge sample of lenses, classifying each
lens in according to the estimated values of the model parameters
\item data from GL can be complemented
with other sources of information from dynamics, fundamental
plane, fit of synthetic stellar population models to lens spectra
to extract the stellar $M/L$ ratio and velocity dispersion, ecc.
and thus to make a direct comparison between stellar and dark
component.
\item it is important not only to determine how much dark matter
exists in galaxies, but also where its contribution becomes
significant, being scale radius $r_0$ a relevant parameter in this
direction
\item at least in the range of redshift from $z=0$ till to $z=1$
we could analyze the evolution with z of estimated lens
parameters, M/L ratio, velocity dispersion predicted by model,
size of galactic haloes and finally the behaviour of the dark
matter content.
\end{itemize}

\end{document}